# Capacity Bounds for Multiuser Channels with Non-Causal Channel State Information at the Transmitters

Reza Khosravi-Farsani, Farokh Marvasti

Adnaced Communications Research Institue (ACRI)
Department of Electrical Engineering, Sharif University of Technology, Tehran, Iran
Email: reza_khosravi@alum.sharif.ir, marvasti@sharif.ir

*Abstract*—In this paper, capacity inner and outer bounds are established for multiuser channels with Channel State Information (CSI) known non-causally at the transmitters: The Multiple Access Channel (MAC), the Broadcast Channel (BC) with common information, and the Relay Channel (RC). For each channel, the actual capacity region is also derived in some special cases. Specifically, it is shown that for some deterministic models with non-causal CSI at the transmitters, similar to Costa's Gaussian channel, the availability of CSI at the deterministic receivers does not affect the capacity region.

## I. INTRODUCTION

Channels with Channel State Information (CSI) known at the transmitters have been investigated in many papers. This study was initiated by Shannon [1] where he considered a single-user channel with CSI known causally at the transmitter and derived its capacity. The capacity of the scenario where the transmitter has access to CSI non-causally was established in [2]. In [3] Costa considered a Gaussian channel with additive interference which is known non-causally at the transmitter and modeled as CSI. He showed that for this channel the capacity is the same as when the receiver is also informed of CSI, thereby, a full interference cancellation is achieved and the capacity is similar to the case of no interference. Up to now, many researchers have attempted to establish capacity results for multiuser channels with CSI. Specifically, Shannon's result for the case of causal CSI [1] was extended to some multiuser scenarios [4], [5] wherein capacity theorems were also established. Moreover, in [6] it was shown that Costa's result [3] regarding achieving full interference cancellation when the interference is known non-causally at the transmitter, can be extended to some multiuser channels. For the case of non-causal CSI, capacity inner and upper bounds have been also established for the discrete multiuser channels [4], [7], [8]; however, there exist only a few cases for which a full characterization of the capacity region has been derived. For a comprehensive review of the existent results regarding capacity bounds for different communication scenarios with CSI, see [9], [10] and the literature therein.

In this paper, we show that for the deterministic single-user channel with CSI known non-causally at the transmitter, similar to Costa's Gaussian channel, the capacity is the same as when the receiver is also informed of CSI. Then, we study capacity bounds for multiuser channels with non-causal CSI. Firstly, we consider the Multiple Access Channel (MAC) with asymmetric and correlated CSI at the transmitters. We establish the capacity region for the deterministic orthogonal MAC with correlated CSI at the transmitters, and demonstrate that in this case the capacity region is the same as when the receiver is also informed of CSI. Also, we derive a new outer bound on the capacity region of the (general) MAC with correlated CSI at the transmitters.

As a second multiuser scenario, we consider the two-user Broadcast Channel (BC) with common message and with CSI known non-causally at the transmitter. We first derive a new inner bound on the capacity region of this channel. The main framework of our achievability scheme is based on applying a (multivariate) random binning technique. We also prove that our achievable rate region *strictly* contains previously derived ones in [11, Sec. V] and [12, p. 7-53] for the channel, as a subset.

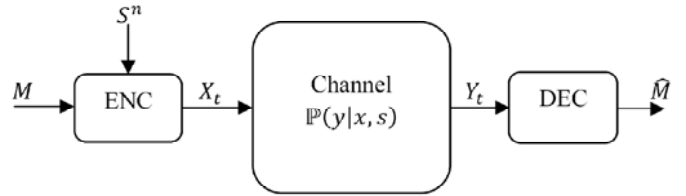

Figure 1. The single-user channel with non-causal CSI.

Moreover, we derive an outer bound for the capacity region of the channel. When there is no CSI, our outer bound reduces to that one derived by Nair and El Gamal in [13] for the BC without CSI. Then, we establish new capacity theorems for the channel in the following cases: 1) Two classes of deterministic BCs with CSI known only at the transmitter, 2) The semi-deterministic BC with CSI known at the transmitter and also at the non-deterministic receiver, 3) The more-capable BC with CSI known at the transmitter and both receivers, 4) A special case of the degraded BC with CSI known only at the transmitter, wherein the non-degraded receiver receives a deterministic function of channel input and channel state. Moreover, in those cases where a receiver receives a deterministic function of channel input and channel state, we show that assuming this receiver to be informed of CSI does not affect the capacity region.

Finally, we study the Relay Channel (RC) with CSI. We first consider the case where both the transmitter and the relay have access to perfect CSI non-causally. We derive a new achievable rate for this channel using the partial decode-and-forward technique. Then, we consider the case of degraded CSI wherein CSI at the relay is a degraded version of CSI at the transmitter. We derive an achievable rate for this scenario using the decode-and-forward technique. As the last scenario, we study the degraded Gaussian RC with asymmetric additive interferences. We consider an interesting scenario regarding the availability of the interferences at the users, as follows. The CSI is composed of two parts: One part is the interference added to the relay received signal, and the other is the interference added to the received signal at the receiver (destination). We assume that the transmitter has access non-causally to both interferences, the relay has access non-causally to the part of interference which is added to the received signal at the receiver, and the receiver has access to the part of interference which is added to the received signal at the relay. In this scenario, no interference subtraction can be performed at the relay and the receiver; nonetheless, we show that a full interference cancellation can be achieved, which yields the capacity of the channel.

The main results are stated in Section II. In Appendix, we have provided an outline of the proofs for most of our theorems. The proof of the remaining results will be reported in [14].

## II. MAIN RESULTS

In this section, we state the main results of the paper. Notations are as follows: Random Variables (RV) and their realization are denoted by upper case and lower case letters, respectively. For a RV $X$ with the range set $\mathcal{X}$, the Probability Distribution Function (PDF) is represented by $P_X(x)$, where $x \in X$. A Gaussian distributed RV, e.g., $X$, with zero mean and variance $\rho^2$ is denoted by $X \sim \mathcal{N}(0, \rho^2)$.

## A. The single-user channel with CSI

*Definition:* The single-user channel with CSI, denoted by $\{S, P_S(s), \mathcal{X}, \mathcal{Y}, \mathbb{P}(y|x,s)\}$, is a channel with input alphabet $\mathcal{X}$ and output alphabet $\mathcal{Y}$. The RV $S$ denotes the channel state which is distributed over the alphabet $S$ according to the known PDF $P_S(s)$. For discrete channels all alphabets are finite sets. The transition probability function $\mathbb{P}(y|x,s)$ describes the relation of channel input, channel state and channel output. The channel model has been depicted in Fig. 1. In this paper, we assume that the state process is non-causally known at the transmitter. The definition of the code and also the capacity for the channel can be found in the literature [9]. Here, due to limited space, they will be omitted.

As mentioned, the capacity of this channel was derived in [2] which is given by the following:

$$\max_{P_{XU|S}} I(U;Y) - I(U;S) \tag{1}$$

Also, for the case where the CSI is available at both transmitter and receiver, the capacity is expressed as [15]:

$$\max_{P_{X|S}} I(X;Y|S) \tag{2}$$

Now, let us assume that the channel is deterministic: There exists a deterministic function $f: \mathcal{X} \times S \to \mathcal{Y}$ such that $Y = f(X,S)$. For this model, we show that similar to Costa's Gaussian channel with additive interference [3], the capacity when CSI is only available at the transmitter is the same as when it is available at both transmitter and receiver.

***Observation:*** *The capacity of the deterministic single-user channel with non-causal CSI at the transmitter is similar to the case where CSI is available at both transmitter and receiver and is given by:*

$$\max_{P_{X|S}} H(Y|S) \tag{3}$$

*Proof:* The achievability is derived from (1) by setting $U = Y$. On the one hand, for deterministic channel, (2) reduces to (3). ∎

As the channel is deterministic (noiseless), this scenario may be well named as "***clean writing on dirty paper***". Subsequently, we demonstrate a same result for some multiuser channels with non-causal CSI at the transmitters.

## B. The MAC with CSI

Now, consider the MAC with non-causal CSI at the transmitters, as depicted in Fig. 2.

*Definition:* The two-user MAC with CSI, denoted by $\{S_1, S_2, P_{S_1S_2}(s_1,s_2), \mathcal{X}_1, \mathcal{X}_2, \mathcal{Y}, \mathbb{P}(y|x_1,x_2,s_1,s_2)\}$, is a channel with input alphabets $\mathcal{X}_1, \mathcal{X}_2$ and output alphabet $\mathcal{Y}$. The channel state is the random pair $(S_1, S_2)$ which is distributed over the set $S_1 \times S_2$ according to $P_{S_1S_2}(s_1,s_2)$. The transition probability function $\mathbb{P}(y|x_1,x_2,s_1,s_2)$ describes the relation of channel inputs, channel state and channel output. As depicted in Fig. 2., the first transmitter has access to $S_1$ and the second one has access to $S_2$, both non-causally. Each transmitter sends a private message over the channel and the receiver decodes both messages.

The MAC with CSI has been investigated in [5], [7], [16]; specifically, for the case of correlated $S_1, S_2$, an achievable rate region has been reported for the channel in [16]:

***Lemma 1*** **[16]:** *The following rate region is achievable for the MAC with non-causal CSI at the transmitters:*

$$\bigcup_{P_{X_1V_1|S_1}P_{X_2V_2|S_2}} \begin{cases} (R_1, R_2) \in \mathbb{R}_+^2: \\ R_1 \leq I(V_1; Y|V_2) - I(V_1; S_1|V_2) \\ R_2 \leq I(V_2; Y|V_1) - I(V_2; S_2|V_1) \\ R_1 + R_2 \leq I(V_1, V_2; Y) - I(V_1, V_2; S_1, S_2) \end{cases} \tag{4}$$

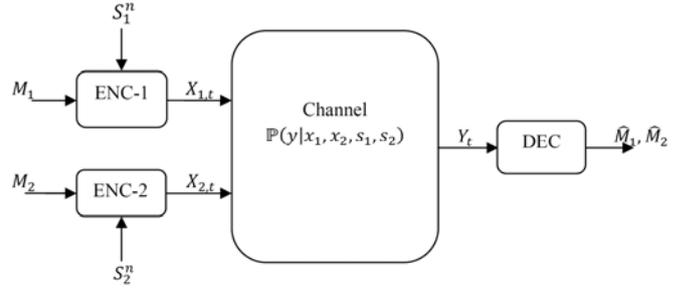

Figure 2. The MAC with asymmetric non-causal CSI.

In [6], it was shown that for Gaussian MAC with additive interference known non-causally at both transmitters the capacity region is the same as when the interference is known also at the receiver. Now the question is that if this property holds for the deterministic MAC, similar to the single-user channel? Unfortunately, using the achievable rate region (4), the capacity region of the deterministic MAC with non-causal CSI at the transmitters cannot be derived in general. Due to this, we restrict our attention to a subclass of MACs called *orthogonal MAC*.

*Definition:* The two-user MAC with CSI is said to be orthogonal if the receiver alphabet $\mathcal{Y} = \mathcal{Y}_1 \times \mathcal{Y}_2$ and the channel transition probability function satisfies:

$$\mathbb{P}(y|x_1,x_2,s_1,s_2) = \mathbb{P}(y_1|x_1,s_1)\mathbb{P}(y_2|x_2,s_2) \tag{5}$$

For the orthogonal MAC with non-causal CSI at the transmitters, when $S_1, S_2$ are independent, one can show that the capacity region is given by:

$$\bigcup_{P_{X_1V_1|S_1}P_{X_2V_2|S_2}} \begin{cases} (R_1, R_2) \in \mathbb{R}_+^2: \\ R_1 \leq I(V_1; Y_1) - I(V_1; S_1) \\ R_2 \leq I(V_2; Y_2) - I(V_2; S_2) \end{cases} \tag{6}$$

For the case of correlated $S_1, S_2$, the channel is not decomposed in two separate ones and hence establishing the capacity region in general case is still difficult. Nevertheless, in the following, we derive the capacity region of the deterministic orthogonal MAC with CSI for the case of correlated $S_1, S_2$. In this model, there exist deterministic functions $f_1: \mathcal{X}_1 \times S_1 \to \mathcal{Y}_1$ and $f_2: \mathcal{X}_2 \times S_2 \to \mathcal{Y}_2$ such that $Y_1 = f_1(X_1, S_1)$ and $Y_2 = f_2(X_2, S_2)$.

***Proposition 1:*** *The capacity region of the deterministic orthogonal MAC with non-causal CSI at the transmitters, where $S_1, S_2$ are correlated, is the same as when both $S_1, S_2$ are also available at the receiver and is given by:*

$$\bigcup_{P_{X_1|S_1}P_{X_2|S_2}} \begin{cases} (R_1, R_2) \in \mathbb{R}_+^2: \\ R_1 \leq H(Y_1|S_1), R_2 \leq H(Y_2|S_2) \end{cases} \tag{7}$$

*Proof:* Refer to Appendix. ∎

To the best of our knowledge, there is no capacity outer bound for the general MAC with non-causal CSI at the transmitters. In the next theorem, we present an outer bound on the capacity region of this channel.

***Theorem 1:*** *The following rate region is an outer bound for the capacity region of the two-user MAC with non-causal CSI at the transmitters ($S_1, S_2$ are correlated):*

$$\bigcup_{P_{X_1X_2V_1V_2|S_1S_2}} \begin{cases} (R_1, R_2) \in \mathbb{R}_+^2: \\ R_1 \leq I(V_1; Y|V_2) + I(V_2; S_2) - I(V_1, V_2; S_1, S_2) \\ R_2 \leq I(V_2; Y|V_1) + I(V_1; S_1) - I(V_1, V_2; S_1, S_2) \\ R_1 + R_2 \leq I(V_1, V_2; Y) - I(V_1, V_2; S_1, S_2) \end{cases} \tag{8}$$

*Proof:* Refer to Appendix. ∎

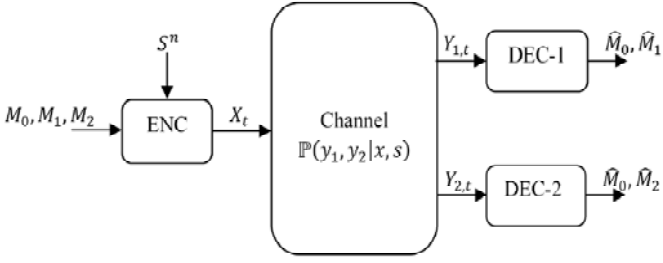

Figure 3. The BC with non-causal CSI.

*Remark:* Consider the following rate region:

$$\bigcup_{P_{X_1 X_2 V_1 V_2 | S_1 S_2}} \begin{Bmatrix} (R_1, R_2) \in \mathbb{R}_+^2: \\ R_1 \leq I(V_1; Y | V_2) - I(V_1; S_1 | V_2) \\ R_2 \leq I(V_2; Y | V_1) - I(V_2; S_2 | V_1) \\ R_1 + R_2 \leq I(V_1, V_2; Y) - I(V_1, V_2; S_1, S_2) \end{Bmatrix} \quad (9)$$

The mutual information functions in the rate region (9) are identical to the achievable rate region (4). However, the union in (9) is taken over a larger set of joint PDFs. It can be easily seen that the derived outer bound (8) is a subset of the rate region (9). Therefore, the rate region (9) also constitutes an outer bound (weaker than (8)) on the capacity region.

### C. The BC with CSI

Then, consider the BC with common message and with CSI known non-causally at the transmitter.

*Definition:* The two-user BC with CSI, denoted by $\{S, P_S(s), \mathcal{X}, \mathcal{Y}_1, \mathcal{Y}_2, \mathbb{P}(y_1, y_2 | x, s)\}$, is a channel with input alphabet $\mathcal{X}$ and output alphabets $\mathcal{Y}_1, \mathcal{Y}_2$. The channel state is the RV $S$ which is distributed over the set $\mathcal{S}$ according to $P_S(s)$. The transition probability function $\mathbb{P}(y_1, y_2 | x, s)$ describes the relation of channel input, channel state and channel outputs. The transmitter has access to CSI non-causally. The channel model has been depicted in Fig. 3. The transmitter sends two private messages, as well as a common message over the channel. Each receiver decodes its respective private message and also the common message.

Firstly, we derive a new achievable rate region for this channel.

*Theorem 2:* The following rate region is achievable for the two-user BC with common message and with CSI known non-causally at the transmitter:

$$\bigcup_{P_{XVUW|S}} \begin{Bmatrix} (R_0, R_1, R_2) \in \mathbb{R}_+^3: \\ R_0 + R_1 \leq I(W, V; Y_1) - I(W, V; S) \\ R_0 + R_2 \leq I(W, U; Y_2) - I(W, U; S) \\ R_0 + R_1 + R_2 \leq I(W, V; Y_1) + I(U; Y_2 | W) \\ \qquad - I(V; U | W) - I(V, U, W; S) \\ R_0 + R_1 + R_2 \leq I(V; Y_1 | W) + I(W, U; Y_2) \\ \qquad - I(V; U | W) - I(V, U, W; S) \\ 2R_0 + R_1 + R_2 \leq I(W, V; Y_1) + I(W, U; Y_2) \\ \qquad - I(V; U | W) - I(V, U, W; S) - I(W; S) \end{Bmatrix}$$

(10)

*Proof:* Refer to [14]. ∎

*Remarks:*

1. By setting $S \equiv \emptyset$, the achievable rate region (10) reduces to Marton's achievable rate region for the two-user BC [17], [18].

2. Our achievable rate region (10) *strictly* contains both those ones derived in [11, Sec. V] and [12, p. 7-53], as a subset. The proof can be found in Appendix.

The achievable rate region (10) is optimal in some special cases. Specifically, consider the deterministic BC with CSI known non-causally at the transmitter: There exist deterministic functions $f_1: \mathcal{X} \times \mathcal{S} \to \mathcal{Y}_1$ and $f_2: \mathcal{X} \times \mathcal{S} \to \mathcal{Y}_2$ such that $Y_1 = f_1(X, S)$ and $Y_2 = f_2(X, S)$. In [6] and [11], it was shown that the capacity region of the two-user Gaussian BC (without common message) with additive interferences is the same as when the interferences (CSI) are also known at both receivers. In the next theorem we derive a similar result for the deterministic BC with CSI.

*Theorem 3:* The capacity region of the deterministic BC (without common message) with CSI known non-causally at the transmitter is the same as when CSI is also known at both receivers and is given by:

$$\bigcup_{P_{X|S}} \begin{Bmatrix} (R_1, R_2) \in \mathbb{R}_+^2: \\ R_1 \leq H(Y_1 | S) \\ R_2 \leq H(Y_2 | S) \\ R_1 + R_2 \leq H(Y_1, Y_2 | S) \end{Bmatrix} \quad (11)$$

*Proof:* The achievability is obtained from (10) by setting $R_0 = 0$, $W \equiv \emptyset$, $V = Y_1$ and $U = Y_2$. The converse part is derived using the cut-set outer bound for channels with CSI [14]. ∎

*Remark:* This is the first class of BCs with CSI known non-causally only at the transmitter for which the capacity region is characterized.

For the case of transmitting both common and private messages, we derive the capacity region of the deterministic BC with CSI satisfying:

$$I(Y_1; Y_2 | S) = 0 \quad (12)$$

for all joint PDFs $P_{X|S}$.

*Proposition 2:* The capacity region of the deterministic BC with common message and with CSI known non-causally only at the transmitter which satisfies the condition (12) is given by:

$$\bigcup_{P_{X|S}} \begin{Bmatrix} (R_1, R_2) \in \mathbb{R}_+^2: \\ R_0 + R_1 \leq H(Y_1 | S) \\ R_0 + R_2 \leq H(Y_2 | S) \end{Bmatrix} \quad (13)$$

*Proof:* Refer to Appendix.

*Remarks:*

1. Proposition 2 generalizes the result of [19, Sec. IV.A] to the case of both common and private messages.

2. In general, the capacity result established in Proposition 2, cannot be deduced by the achievable rate region previously derived in [11, Sec. V].

We also derive an outer bound on the capacity region of the channel.

*Theorem 4:* The following rate region is an outer bound for the two-user BC with non-causal CSI at the transmitter:

$$\bigcup_{P_{XUV|S}} \begin{Bmatrix} (R_0, R_1, R_2) \in \mathbb{R}_+^3: \\ R_0 + R_1 \leq I(V; Y_1 | S) \\ R_0 + R_2 \leq I(U; Y_2 | S) \\ R_0 + R_1 + R_2 \leq I(X; Y_1 | U, S) + I(U; Y_2 | S) \\ R_0 + R_1 + R_2 \leq I(X; Y_2 | V, S) + I(V; Y_1 | S) \end{Bmatrix}$$

(14)

*Proof:* Refer to Appendix. ∎

*Remarks:*

1. The rate region (14) continues to be an outer bound for the channel when CSI is also available at both receivers.

2. By setting $S \equiv \emptyset$ the rate region (14) reduces to Nair-El Gamal outer bound [13] on the capacity region of the two-user BC.

Now, we present some classes of channels for which the achievability scheme (10) and the outer bound (14) agree, which yields the capacity region. As the first scenario, in the next theorem, we establish the capacity region of the two-user BC with non-causal CSI at the transmitter where one receiver receives a deterministic function of the channel input and the channel state, and the other has access to CSI.

**Theorem 5:** *Consider the semi-deterministic BC (without common message) with CSI known non-causally at the transmitter, wherein there exists a deterministic function $f_1: \mathcal{X} \times \mathcal{S} \to \mathcal{Y}_1$ such that $Y_1 = f_1(X, S)$. Moreover, assume that the CSI is also known at the second receiver. In this case, the capacity region is the same as when both receivers have access to CSI and is given by:*

$$\bigcup_{P_{XU|S}} \begin{cases} (R_1, R_2) \in \mathbb{R}_+^2: \\ R_1 \leq H(Y_1|S) \\ R_2 \leq I(U; Y_2|S) \\ R_1 + R_2 \leq H(Y_1|U, S) + I(U; Y_2|S) \end{cases}$$

(15)

*Proof:* The achievability is obtained from (10) by setting $W \equiv \emptyset$, $V = Y_1$ and replacing $Y_2$ with $(Y_2, S)$. The converse part is derived from (14) by considering that $I(V; Y_1|S) \leq H(Y_1|S)$, and $Y_1 = f_1(X, S)$. ∎

As another scenario, we also determine the capacity region of the more-capable BC with non-causal CSI at the transmitter, for the case where both receivers have access to CSI.

*Definition:* The BC with CSI known non-causally at the transmitter is said to be more-capable if

$$I(X; Y_2|S) \leq I(X; Y_1|S) \quad (16)$$

for all joint PDFs $P_{X|S}(x|s)$.

**Proposition 3:** *The capacity region of the two-user more-capable BC (16) with common message and with CSI known non-causally at the transmitter and also at both receivers is given by:*

$$\bigcup_{P_{XU|S}} \begin{cases} (R_0, R_1, R_2) \in \mathbb{R}_+^3: \\ R_0 + R_2 \leq I(U; Y_2|S) \\ R_0 + R_1 + R_2 \leq I(X; Y_1|U, S) + I(U; Y_2|S) \\ R_0 + R_1 + R_2 \leq I(X; Y_1|S) \end{cases}$$

(17)

*Proof:* The achievability is derived from (10) by setting $W \equiv U$, $V \equiv X$, and also replacing $Y_1$ with $(Y_1, S)$ and $Y_2$ with $(Y_2, S)$. The converse part is obtained from the outer bound (14) where the relation (16) is also exploited. ∎

In [4] an achievable rate region was derived for the degraded BC with non-causal CSI at the transmitter using superposition coding technique. However, the capacity region remains still unknown. In the following, we derive the capacity region of this channel under the condition that the signal of the stronger receiver is a deterministic function of the channel input and the channel state.

*Definition:* The two-user BC with CSI is said to be degraded if

$$\mathbb{P}(y_1, y_2|x, s) = \mathbb{P}(y_1|x, s)\mathbb{P}(y_2|y_1) \quad (18)$$

It can be easily verified that the degraded BC with CSI (18), also satisfies the more-capable condition (16).

**Theorem 6:** *The capacity region of the degraded BC (18) with common message and with non-causal CSI at the transmitter, wherein there exists a deterministic function $f_1: \mathcal{X} \times \mathcal{S} \to \mathcal{Y}_1$ such that $Y_1 = f_1(X, S)$, is given by the following:*

$$\bigcup_{P_{XU|S}} \begin{cases} (R_0, R_1, R_2) \in \mathbb{R}_+^3: \\ R_1 \leq H(Y_1|U, S) \\ R_0 + R_2 \leq I(U; Y_2) - I(U; S) \end{cases}$$

(19)

*Moreover, the availability of CSI at the stronger receiver, i.e., $Y_1$, does not affect the capacity region.*

*Proof:* Refer to Appendix. ∎

### D. The RC with CSI

Finally, consider the RC with non-causal CSI.

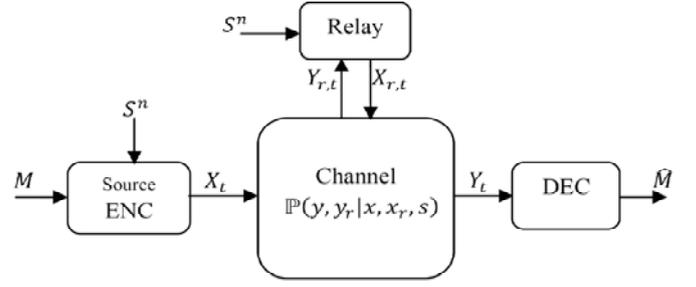

Figure 4. The RC with non-causal CSI.

*Definition:* The RC with CSI denoted by $\{\mathcal{S}, P_S(s), \mathcal{X}, \mathcal{X}_r, \mathcal{Y}_r, \mathcal{Y}, \mathbb{P}(y, y_r|x, x_r, s)\}$, is a channel with the source input alphabet $\mathcal{X}$, the relay input alphabet $\mathcal{X}_r$, the relay output alphabet $\mathcal{Y}_r$ and the receiver (destination) output alphabet $\mathcal{Y}$. The channel state is the RV $S$ which is distributed over the set $\mathcal{S}$ according to $P_S(s)$. The transition probability function of the channel $\mathbb{P}(y, y_r|x, x_r, s)$ describes the relation of channel input, channel state and channel outputs. The channel model has been depicted in Fig. 4. The transmitter sends a message over the channel and the relay assists the transmission of information to the receiver. The receiver is required to decode the transmitted message.

In [6], the case where perfect CSI is known non-causally at the transmitter and also at the relay node was considered and an achievable rate was derived for the channel using the decode-and-forward technique. In the next theorem, we establish a new achievable rate for this channel based on the partial decode-and-forward technique [20].

**Proposition 4:** *The following is an achievable rate for the RC with CSI known non-causally at the transmitter and at the relay node:*

$$\sup \min \begin{cases} I(V, U, U_r; Y) - I(V, U, U_r; S), \\ I(V, U; Y|U_r) + I(U; Y_r|U_r, S) - I(V, U; S|U_r), \\ I(V; Y|U, U_r) + I(U; Y_r|U_r, S) - I(V; S|U, U_r) \end{cases}$$

(20)

*where the supremum in (20) is taken over all joint PDFs $P_{X_r U_r X V U|S}(x_r, u_r, x, v, u|s)$ that factor as:*

$$P_{X_r U_r X V U|S} = P_{X_r U_r|S} P_{XVU|U_r S} \quad (21)$$

*and subjected to:*

$$\begin{cases} I(V, U; Y|U_r) > I(V, U; S|U_r) \\ I(V; Y|U, U_r) > I(V; S|U, U_r) \end{cases} \quad (22)$$

*Proof:* Refer to [14]. ∎

Note that the capacity of the deterministic and orthogonal RC without CSI is obtained using the achievable rate derived based on the partial decode-and-forward technique [20], [21], [22]; unfortunately, for the channel with CSI it does not seem that the achievable rate (20) leads to the capacity for the deterministic or orthogonal RC with CSI.

As the second scenario, we consider the RC with CSI where CSI is a correlated pair $S = (S_1, S_2)$: The transmitter has access non-causally to both $S_1$ and $S_2$, while the relay has access non-causally only to $S_1$. In the next theorem, we derive an achievable rate for this channel using the decode-and-forward technique.

**Theorem 7:** *Consider the RC with CSI $S = (S_1, S_2)$ wherein the transmitter has access non-causally to both $S_1$ and $S_2$, but the relay has access non-causally only to $S_1$. The following is an achievable rate:*

$$\sup \min \begin{cases} I(U, U_r; Y) - I(U, U_r; S_1, S_2), \\ I(U; Y_r|U_r, S_1) - I(U; S_2|U_r, S_1) \end{cases}$$

(23)

*where the supremum in (23) is taken over all joint PDFs $P_{X_r U_r X U|S_1 S_2}(x_r, u_r, x, u|s_1, s_2)$ that factor as:*

$$P_{X_r U_r X U|S_1 S_2} = P_{X_r U_r|S_1} P_{XU|U_r S_1 S_2} \quad (24)$$

*Proof:* Refer to [14]. ∎

*Remark:* The achievable rate (23) will be used in Theorem 8 to prove a capacity result for the Gaussian channel with additive interferences.

Finally, consider the degraded Gaussian RC with additive interferences. The channel is formulated as follows:

$$\begin{cases} Y_r = X + S_r + Z_r \\ Y = X + X_r + S_d + Z_r + Z_d \end{cases} \quad (25)$$

where $X$ and $X_r$ are the transmitter input and the relay input, respectively; $Y_r$ and $Y$ are the received signals at the relay and the receiver, respectively; $Z_r \sim \mathcal{N}(0, N_r)$ and $Z_d \sim \mathcal{N}(0, N_d)$ are independent Gaussian noises and $S_r \sim \mathcal{N}(0, P_{S_r})$ and $S_d \sim \mathcal{N}(0, P_{S_d})$ are additive Gaussian interferences. The Gaussian noises $Z_r$ and $Z_d$ are independent of $(S_r, S_d)$, but $S_r$ and $S_d$ can be correlated. The input signals are power constrained, i.e., $\mathbb{E}[X^2] \leq P$ and $\mathbb{E}[X_r^2] \leq P_r$.

Note that the channel (25) when $S_r \equiv S_d \equiv \emptyset$ reduces to the degraded Gaussian RC for which the capacity was established in [20]. This channel was considered in [6] for the case of $S_r \equiv S_d \equiv \tilde{S}$, where $\tilde{S}$ is non-causally known at the transmitter and at the relay. It was shown that in this case the capacity is the same as when the interference $\tilde{S}$ is also available at the receiver; in other words, the capacity is the same as when there is no additive interference in the channel. In the following, by considering the generalized model (25) we derive new results.

Firstly, we consider the case where the relay node knows (non-causally) the pair $(S_r, S_d)$, but the transmitter knows only the part $S_d$. Since the relay knows $S_r$, this part of interference can be subtracted from the received signal at the relay; thereby the channel is equivalent to the following:

$$\begin{cases} Y_r = X + Z_r \\ Y = X + X_r + S_d + Z_r + Z_d \end{cases} \quad (26)$$

Again, because the relay knows $S_d$, it can add this part of the interference to its received signal; thereby, the channel is equivalent to the following:

$$\begin{cases} Y_r = X + S_d + Z_r \\ Y = X + X_r + S_d + Z_r + Z_d \end{cases} \quad (27)$$

where $S_d$ is available non-causally at the transmitter and the relay node. On the other hand, the model (27) is similar to that one considered in [6]. Therefore, the following result is established.

**Proposition 5:** *The capacity of the degraded Gaussian RC with additive interferences (25) in which the relay has access non-causally to the pair $(S_r, S_d)$, but the transmitter has access non-causally only to the part $S_d$, is the same as when there is no additive interference in the channel and is given by:*

$$\sup_{0 \leq \alpha \leq 1} \min \left\{ C\left(\frac{P + P_r + 2\sqrt{\bar{\alpha} P P_r}}{N_r + N_d}\right), C\left(\frac{\alpha P}{N_r}\right) \right\} \quad (28)$$

where $C(x) \triangleq \frac{1}{2} \log(1 + x)$.

In the previous scenario to achieve a full interference cancellation it was required that the relay knows both interferences, i.e., $(S_r, S_d)$. Now, let us consider an interesting scenario regarding the availability of the interferences at the users, as follows. We assume that the transmitter has access non-causally to both interferences $(S_r, S_d)$, the relay has access non-causally to the part of interference which is added to the received signal at the receiver, i.e., $S_d$, and the receiver has access to the part of interference which is added to the received signal at the relay, i.e., $S_r$. Note that in this scenario no interference subtraction can be performed at the relay and the receiver; nonetheless, in the following theorem we show that a full interference cancellation can be achieved.

**Theorem 8:** *The capacity of the degraded Gaussian RC with additive interferences (25) wherein the transmitter knows non-causally both interferences $S_r, S_d$, the relay knows non-causally $S_d$ and the receiver knows $S_r$, is the same as when there is no interference in the channel and is given by (28).*

*Proof:* Refer to Appendix. ∎

## III. CONCLUSION

In this paper, new capacity inner and outer bounds were established for the multiuser channels with non-causal CSI at the transmitters, and also the actual capacity region was derived in some special cases. It seems that for discrete deterministic channels with non-causal CSI at the transmitters, the availability of CSI at the deterministic receivers does not affect the capacity region. This result is similar to Gaussian channel with additive interference wherein the capacity when the transmitter has access to CSI non-causally is the same as when the receiver also knows it.

## REFERENCES

[1] C. E. Shannon, "Channels with side information at the transmitter," *IBM Journal Res. and Develop.*, vol. 2, pp. 289-293, Oct. 1958.

[2] S. I. Gel'fand and M. S. Pinsker, "Coding for channel with random parameters," *Prob. of Contr. Inf. Theory*, vol. 9, no. 1, pp. 19–31, 1980.

[3] M. H. M. Costa, "Writing on dirty paper," *IEEE Trans. Inf. Theory*, vol. 29, no. 3, pp. 439–441, May 1983.

[4] Y. Steinberg, "Coding for the degraded broadcast channel with random parameters, with causal and noncausal side information," *IEEE Trans. Inf. Theory*, vol. 51, no. 8, pp. 2867–2877, Aug. 2005.

[5] S. Sigurjonsson and Y. H. Kim, "On multiple user channels with state information at the transmitters," in *Proc. IEEE Int. Symp. Information Theory (ISIT'05)*, Adelaide, Australia, Sep. 2005, pp. 72-76.

[6] Y-H. Kim, A. Sutivong, and S. Sigurjonsson, "Multiple user writing on dirty paper," in *Proc. IEEE Int. Symp. Information Theory (ISIT'04)*, Chicago, USA, Jun./Jul 2004, p. 534.

[7] S. A. Jafar, "Capacity with causal and noncausal side information—A unified view," *IEEE Trans. Inf. Theory*, vol. 52, no. 12, pp. 5468–5475, Dec. 2006.

[8] A. Zaidi, S. P. Kotagiri, J. N. Laneman, and L. Vandendorpe, "Cooperative relaying with state available noncausally at the relay", *IEEE Trans. Inf. Theory*, vol. 56, no. 5, pp. 2272–2298, May 2010.

[9] G. Keshet, Y. Steinberg, and N. Merhav, "Channel coding in the presence of side information," *Foundation and Trends in Communications and Information Theory*, 2008.

[10] R. Khosravi-Farsani, F. Marvasti, "Multiple access channels with cooperative encoders and channel state information" *Submitted to European Transactions on Telecommunications.* Sep. 2010. Available at: http://arxiv.org/abs/1009.6008.

[11] Y. Steinberg and S. Shamai (Shitz), "Achievable rates for the broadcast channel with states known at the transmitter," in *Proc. IEEE Int. Symp. Information Theory (ISIT'05)*, Adelaide, Australia, Sep. 2005, pp. 2184-2188.

[12] A. El Gamal and Y. H. Kim, "Lectures on network information theory," 2010. Available online at: http://arxiv.org/abs/1001.3404.

[13] C. Nair and A. El Gamal, "An outer bound to the capacity region of the broadcast channel," *IEEE Trans. Info. Theory*, vol. IT-53, pp. 350–355, Jan. 2007.

[14] R. Khosravi-Farsani, F. Marvasti, "Capacity bounds for multiuser channels with channel state information", in preparation.

[15] T. M. Cover and M. Chiang, "Duality between channel capacity and rate distortion with two-sided state information," *IEEE Trans. Inf. Theory*, vol. 48, no. 6, pp. 1629–1638, Jun. 2002.

[16] T. Philosof and R. Zamir, "On the loss of single-letter characterization: The dirty multiple access channel," *IEEE Trans. Inf. Theory*, vol. 55, no. 6, pp. 2442–2454, Jun. 2009.

[17] K. Marton, "A coding theorem for the discrete memoryless broadcast channel," *IEEE Trans. Inf. Theory*, vol. 25, no. 3, pp. 306–311, May 1979.

[18] Y. Liang and G. Kramer, "Rate regions for relay broadcast channels," *IEEE Trans. Inf. Theory*, vol. 53, no. 10, pp. 3517–3535, Oct. 2007.

[19] C. Nair, A. El Gamal, and Y-K. Chia, "An achievability scheme for the compound channel with state noncausally available at the encoder," available at: http://arxiv.org/abs/1004.3427.

[20] T. Cover and A. El Gamal, "Capacity theorems for the relay channel," *IEEE Trans. Inf. Theory*, vol. IT-25, no. 5, pp. 572–584, Sep. 1979.

[21] A. El Gamal and M. Aref, "The capacity of the semideterministic relay channel," *IEEE Trans. Inf. Theory*, vol. IT-28, no. 3, p. 536, May 1982.

[22] A. El Gamal and S. Zahedi, "Capacity of relay channels with orthogonal components," *IEEE Trans. Inf. Theory*, vol. 51, no. 5, pp. 1815–1817, May 2005.

[23] A. Aminzadeh Gohari, A. El Gamal, and V. Anantharam, "On an outer bound and an inner bound for the general broadcast channel," *Int. Symp. on Info. Theory (ISIT)*, Jul. 2010.

# APPENDIX

➢ *Proof of proposition 1:*

The achievability is obtained from (4) by setting $V_i = Y_i, i = 1,2$, where the equations $H(Y_1|Y_2, S_1) = H(Y_1|S_1)$ and $H(Y_2|Y_1, S_2) = H(Y_2|S_2)$ are applied. For the converse part, let us assume that the CSI $(S_1, S_2)$ is available at the receiver. Consider a length-$n$ code with vanishing average error probability for the channel. By Fano's inequality, one can show:

$$nR_1 \leq \sum_{t=1}^{n} I(X_{1,t}; Y_t | X_{2,t}, S_{1,t}, S_{2,t}) + \epsilon_{1,n}$$

$$\stackrel{(a)}{=} \sum_{t=1}^{n} H(Y_{1,t}, Y_{2,t} | X_{2,t}, S_{1,t}, S_{2,t}) + \epsilon_{1,n}$$

$$\stackrel{(b)}{=} \sum_{t=1}^{n} H(Y_{1,t} | X_{2,t}, S_{1,t}, S_{2,t}) + \epsilon_{1,n}$$

$$\leq \sum_{t=1}^{n} H(Y_{1,t} | S_{1,t}) + \epsilon_{1,n}$$

$$(A.1)$$

where $\epsilon_{1,n} \to 0$ as $n \to \infty$, equality (a) holds because $Y_t = (Y_{1,t}, Y_{2,t})$ is a deterministic function of $(X_{1,t}, X_{2,t}, S_{1,t}, S_{2,t})$, and equality (b) holds because $Y_{2,t}$ is a deterministic function of $(X_{2,t}, S_{2,t})$. Similarly, we have:

$$nR_2 \leq \sum_{t=1}^{n} H(Y_{2,t}|S_{2,t}) + \epsilon_{2,n}$$

$$(A.2)$$

where $\epsilon_{2,n} \to 0$ as $n \to \infty$. Note that the PDF $P(x_{1,t}, x_{2,t}|s_{1,t}, s_{2,t}), t = 1, \ldots, n$, does not factor as $P(x_{1,t}|s_{1,t})P(x_{2,t}|s_{2,t})$; however, due to the orthogonal property of the channel the expressions appeared in $(A.1)$ and $(A.2)$ depend only on the marginal PDFs $P(x_{1,t}|s_{1,t}, s_{2,t}) = P(x_{1,t}|s_{1,t})$ and $P(x_{2,t}|s_{1,t}, s_{2,t}) = P(x_{2,t}|s_{2,t})$, respectively. Therefore, without loss of generality we can consider the bounds $(A.1)$ and $(A.2)$ under the PDFs of the form $P(x_{1,t}|s_{1,t})P(x_{2,t}|s_{2,t}), t = 1, \ldots, n$, as desired. ∎

➢ *Proof of Theorem 1:*

Consider a length-$n$ code with vanishing error probability for the channel. Define new RVs $V_{1,t}, V_{2,t}, t = 1, \ldots, n$, as follows:

$$V_{i,t} \triangleq (Y^{t-1}, M_i, S_{i,t+1}^n), \quad i = 1,2$$

$$(A.3)$$

Using Fano's inequality we can write:

$$nR_1 \leq I(M_1; Y^n, M_2) + n\epsilon_{1,n}$$

$$= \sum_{t=1}^{n} I(M_1; Y_t | Y^{t-1}, M_2) + n\epsilon_{1,n}$$

$$\leq \sum_{t=1}^{n} I(M_1, S_{1,t+1}^n, S_{2,t+1}^n; Y_t | Y^{t-1}, M_2) - \sum_{t=1}^{n} I(S_{1,t+1}^n, S_{2,t+1}^n; Y_t | Y^{t-1}, M_1, M_2) + n\epsilon_{1,n}$$

$$= \sum_{t=1}^{n} I(M_1, S_{1,t+1}^n; Y_t | Y^{t-1}, M_2, S_{2,t+1}^n) + \sum_{t=1}^{n} I(S_{2,t+1}^n; Y_t | Y^{t-1}, M_2) - \sum_{t=1}^{n} I(S_{1,t+1}^n, S_{2,t+1}^n; Y_t | Y^{t-1}, M_1, M_2) + n\epsilon_{1,n}$$

$$(A.4)$$

where $\epsilon_{1,n} \to \infty$ as $n \to \infty$. Now, using the Csiszar-Korner lemma, we have:

$$\sum_{t=1}^{n} I(S_{2,t+1}^n; Y_t | Y^{t-1}, M_2) = \sum_{t=1}^{n} I(Y^{t-1}; S_{2,t} | M_2, S_{2,t+1}^n) = \sum_{t=1}^{n} I(Y^{t-1}, M_2, S_{2,t+1}^n; S_{2,t}) = \sum_{t=1}^{n} I(V_{2,t}; S_{2,t})$$

$$(A.5)$$

and also,

$$\sum_{t=1}^{n} I(S_{1,t+1}^n, S_{2,t+1}^n; Y_t | Y^{t-1}, M_1, M_2) = \sum_{t=1}^{n} I(S_{1,t}, S_{2,t}; Y^{t-1} | S_{1,t+1}^n, S_{2,t+1}^n, M_1, M_2)$$

$$= \sum_{t=1}^{n} I(Y^{t-1}, S_{1,t+1}^n, S_{2,t+1}^n, M_1, M_2; S_{1,t}, S_{2,t})$$

$$= \sum_{t=1}^{n} I(V_{1,t}, V_{2,t}; S_{1,t}, S_{2,t})$$

(A.6)

By substituting $(A.5)$ and $(A.6)$ in $(A.4)$, we obtain:

$$nR_1 \leq \sum_{t=1}^{n} I(V_{1,t}; Y_t | V_{2,t}) + I(V_{2,t}; S_{2,t}) - I(V_{1,t}, V_{2,t}; S_{1,t}, S_{2,t}) + n\epsilon_{1,n}$$

(A.7)

Symmetrically, we can derive:

$$nR_2 \leq \sum_{t=1}^{n} I(V_{2,t}; Y_t | V_{1,t}) + I(V_{1,t}; S_{1,t}) - I(V_{1,t}, V_{2,t}; S_{1,t}, S_{2,t}) + n\epsilon_{2,n}$$

(A.8)

where $\epsilon_{2,n} \to \infty$ as $n \to \infty$. For the sum-rate by following the same lines as single-user channel [2] one can show:

$$n(R_1 + R_2) \leq \sum_{t=1}^{n} I(V_{1,t}, V_{2,t}; Y_t) - I(V_{1,t}, V_{2,t}; S_{1,t}, S_{2,t}) + n(\epsilon_{1,n} + \epsilon_{2,n})$$

(A.9)

Then, by applying a standard time-sharing argument the outer bound (8) is derived. ∎

> ***Proof of Remark 2 of Theorem 2:***

First note that by setting $W \equiv \emptyset$ and $R_0 = 0$ in (10), our achievable rate region reduces to the one reported in [12, p.7-53]. Based on the result of [23, Lemma 1], it is readily derived that our achievable rate region (10) strictly contains that of [12, p. 7-53]. In fact, this strict inclusion holds even for the case of $\|S\| = 1$. Now, consider the rate region reported in [11, Sec. V] which is expressed by the following constraints:

$$R_0 \leq [\min\{I(W; Y_1), I(W; Y_2)\} - I(W; S)]_+ \quad (a)$$
$$R_0 + R_1 \leq I(W, V; Y_1) - I(W, V; S) \quad (b)$$
$$R_0 + R_2 \leq I(W, U; Y_2) - I(W, U; S) \quad (c)$$
$$R_0 + R_1 + R_2 \leq -[\max\{I(W; Y_1), I(W; Y_2)\} - I(W; S)]_+ \quad (d)$$
$$+ I(W, V; Y_1) - I(W, V; S)$$
$$+ I(W, U; Y_2) - I(W, U; S) - I(U; V | W, S)$$

(A.10)

where $[a]_+ = \max\{0, a\}$. The achievable rate region of [11, Sec. V] is the set of all triples $(R_0, R_1, R_2)$ satisfying $(A.10)$ for some joint PDFs $P_{XVUW|S}(x, v, u, w|s)$. Given such a joint PDF $P_{VUWX}$, we first show if $(R_0, R_1, R_2)$ satisfies $(A.10)$, it also belongs to our rate region (10). First note that, for every joint PDF $P_{XVUWS}(x, v, u, w, s)$ we have:

$$I(W, V; S) + I(W, U; S) + I(U; V | W, S) = I(V; U | W) + I(V, U, W; S) + I(W; S)$$

(A.11)

To see this equality, consider the left hand side of $(A.11)$. We can write:

$$I(W, V; S) + I(W, U; S) + I(U; V | W, S) = I(W; S) + I(V; S | W) + I(U; V | W, S) + I(W, U; S)$$
$$= I(W; S) + I(V; S, U | W) + I(W, U; S)$$
$$= I(W; S) + I(V; U | W) + I(V; S | W, U) + I(W, U; S)$$
$$= I(W; S) + I(V; U | W) + I(V, U, W; S)$$
$$= \text{Right Hand side of } (A.11).$$

Also, since $[a]_+ \geq a$, we have:

$$-[\max\{I(W;Y_1), I(W;Y_2)\} - I(W;S)]_+ \leq -\max\{I(W;Y_1), I(W;Y_2)\} + I(W;S)$$
$$= \min\{-I(W;Y_1), -I(W;Y_2)\} + I(W;S)$$

$$(A.12)$$

Moreover, by definition of $[a]_+$, one can easily derive:

$$[\min\{I(W;Y_1), I(W;Y_2)\} - I(W;S)]_+ \leq [\max\{I(W;Y_1), I(W;Y_2)\} - I(W;S)]_+$$

$$(A.13)$$

Next, from $(A.10*d)$ and $(A.12)$, we have:

$$R_0 + R_1 + R_2 \leq -I(W;Y_1) + I(W;S)$$
$$+ I(W,V;Y_1) - I(W,V;S)$$
$$+ I(W,U;Y_2) - I(W,U;S) - I(U;V|W,S)$$
$$\stackrel{(a)}{=} I(V;Y_1|W) + I(W,U;Y_2) - I(V;U|W) - I(V,U,W;S)$$

$$(A.14)$$

where equality (a) is obtained by $(A.11)$. Similarly,

$$R_0 + R_1 + R_2 \leq -I(W;Y_2) + I(W;S)$$
$$+ I(W,V;Y_1) - I(W,V;S)$$
$$+ I(W,U;Y_2) - I(W,U;S) - I(U;V|W,S)$$
$$= I(W,V;Y_1) + I(U;Y_2|W) - I(V;U|W) - I(V,U,W;S)$$

$$(A.15)$$

Finally, by adding the two sides of $(A.10*a)$ and $(A.1*d)$, and also considering $(A.13)$, we obtain:

$$2R_0 + R_1 + R_2 \leq I(W,V;Y_1) - I(W,V;S) + I(W,U;Y_2) - I(W,U;S) - I(U;V|W,S)$$
$$\stackrel{(a)}{=} I(W,V;Y_1) + I(W,U;Y_2) - I(V;U|W) - I(V,U,W;S) - I(W;S)$$

$$(A.16)$$

where (a) is due to $(A.11)$. Therefore, by $(A.10*b)$, $(A.10*c)$, $(A.14)$, $(A.15)$, and $(A.16)$, we derive that $(R_0, R_1, R_2)$ belongs to our rate region (10).

Then, we show that our rate region can strictly contain that of [11, Sec. V]. Consider the scenario of broadcasting only a common message, i.e., $R_1 = R_2 = 0$. In this case, due to $(A.10*a)$ it is readily derived that the maximum common rate which can be achieved by the rate region of [11, Sec. V], denoted by $R_0^{SS}$, at most is as follows:

$$R_0^{SS} \leq \max_{P_{XW|S}} [\min\{I(W;Y_1), I(W;Y_2)\} - I(W;S)]_+ \stackrel{(a)}{=} \max_{P_{XW|S}} (\min\{I(W;Y_1), I(W;Y_2)\} - I(W;S))$$

$$(A.17)$$

where equality (a) holds because by setting $W \equiv \emptyset$, the argument of the maximization given above is zero and hence it can take also non-negative values. Now, consider our achievable rate region (10). By setting $R_1 = R_2 = 0$ in (10), we obtain that the following rate can be achieved:

$$R_0 = \max_{P_{XVUW|S}} \left( \min \left( \begin{array}{c} I(W,V;Y_1) - I(W,V;S), \\ I(W,U;Y_2) - I(W,U;S), \\ I(V;Y_1|W) + I(W,U;Y_2) - I(V;U|W) - I(V,U,W;S), \\ I(W,V;Y_1) + I(U;Y_2|W) - I(V;U|W) - I(V,U,W;S), \\ \frac{1}{2}(I(W,V;Y_1) + I(W,U;Y_2) - I(V;U|W) - I(V,U,W;S) - I(W;S)) \end{array} \right) \right)$$

$$(A.18)$$

On the one hand, by setting $U \equiv V \equiv \emptyset$, $(A.18)$ reduces to $(A.17)$, (note that, for every $a$ and $b$, we have $\min(a,b) \leq \frac{a+b}{2}$); therefore, our lower bound on the common rate is at least as large as $R_0^{SS}$ given by $(A.17)$. Now, we show that indeed the common rate achievable by our rate region (10) can be strictly larger than $R_0^{SS}$. To prove this, we adapt a recent result derived in [19] for broadcasting a common message to two receivers where the transmitter has access to non-causal CSI. Precisely speaking, the authors in [19, Th. 1] established the following achievable rate for this scenario, which we denote by $R_0^{NEGC}$,:

$$R_0^{NEGC} = \max_{P_{XVUW|S}} \left( \min \begin{pmatrix} I(W,V;Y_1) - I(W,V;S), \\ I(W,U;Y_2) - I(W,U;S), \\ \frac{1}{2}\big(I(W,V;Y_1) + I(W,U;Y_2) - I(W,V;S) - I(W,U;S) - I(U;V|W,S)\big) \end{pmatrix} \right)$$

(A.19)

In general, $R_0^{NEGC}$ is larger than both our lower bound given in (A.18) and $R_0^{SS}$ given in (A.17). Nevertheless, we demonstrate that there exists channels for which our lower bound and $R_0^{NEGC}$ coincide and achieve the capacity, while $R_0^{SS}$ is strictly suboptimal. In fact, one example is justly the channel considered in [19, Sec. III]. We do not discuss the details of this example as can be found in [19, Sec. III]. For the mentioned example, the authors proved that the capacity can be achieved by setting $W \equiv \emptyset$, $V \equiv Y_1$ and $U \equiv Y_2$ in (A.19); while $R_0^{SS}$ is strictly suboptimal. On the one hand, by this choice of auxiliary RVs our lower bound given in (A.18) and $R_0^{NEGC}$ coincide, which are equal to:

$$\max_{P_{X|S}} \min \left\{ H(Y_1|S), H(Y_2|S), \frac{1}{2}\big(H(Y_1, Y_2|S)\big) \right\}$$

(A.20)

Therefore, for the example given in [19, Sec. III] our lower bound also achieves the capacity, while $R_0^{SS}$ is strictly suboptimal. This proves the desired result. ∎

➤ *Proof of Proposition 2:*

For the direct part, by setting $W \equiv \emptyset$, $V \equiv Y_1$, $U \equiv Y_2$, in the achievable rate region (10) we derive:

$$\begin{aligned} R_0 + R_1 &\leq H(Y_1|S) \\ R_0 + R_2 &\leq H(Y_2|S) \\ 2R_0 + R_1 + R_2 &\leq H(Y_1, Y_2|S) \end{aligned}$$

(A.21)

One the one hand, by the condition (12) we have: $H(Y_1, Y_2|S) = H(Y_1|S) + H(Y_2|S)$. Therefore, the third constraint of (A.21) is redundant. The converse part is derived using the cut-set outer bound for channels with CSI. ∎

➤ *Proof of Theorem 4:*

Consider a length-$n$ code with vanishing average error probability for the channel. Define new RVs $U_t, V_t, t = 1, \ldots, n$, as follows:

$$\begin{aligned} V_t &\triangleq \big(M_0, M_1, Y_1^{t-1}, S^{t-1}, S_{t+1}^n, Y_{2,t+1}^n\big) \\ U_t &\triangleq \big(M_0, M_2, Y_1^{t-1}, S^{t-1}, S_{t+1}^n, Y_{2,t+1}^n\big) \end{aligned}$$

(A.22)

Using Fano's inequality, one can write:

$$n(R_0 + R_1) \leq I(M_0, M_1; Y_1^n, S^n) + n\epsilon_{1,n}$$

$$= \sum_{t=1}^n I\big(M_0, M_1; Y_{1,t} \big| Y_1^{t-1}, S^n\big) + n\epsilon_{1,n}$$

$$\leq \sum_{t=1}^n I\big(M_0, M_1, Y_1^{t-1}, S^{t-1}, S_{t+1}^n, Y_{2,t+1}^n; Y_{1,t} \big| S_t\big) + n\epsilon_{1,n}$$

$$= \sum_{t=1}^n I\big(V_t; Y_{1,t} \big| S_t\big) + n\epsilon_{1,n}$$

(A.23)

where $\epsilon_{1,n} \to \infty$ as $n \to \infty$. Also, we have:

$$n(R_0 + R_2) \leq I(M_0, M_2; Y_2^n, S^n) + n\epsilon_{2,n}$$

$$= \sum_{t=1}^{n} I(M_0, M_2; Y_{2,t} | Y_{2,t+1}^n, S^n) + n\epsilon_{2,n}$$

$$\leq \sum_{t=1}^{n} I(M_0, M_2, Y_1^{t-1}, S^{t-1}, S_{t+1}^n, Y_{2,t+1}^n; Y_{2,t} | S_t) + n\epsilon_{2,n}$$

$$= \sum_{t=1}^{n} I(U_t; Y_{2,t} | S_t) + n\epsilon_{2,n}$$

(A.24)

where $\epsilon_{2,n} \to \infty$ as $n \to \infty$. For the sum rate, i.e., $R_0 + R_1 + R_2$, we can write:

$$n(R_0 + R_1 + R_2) \leq I(M_1; Y_1^n, M_0, M_2, S^n) + I(M_0, M_2; Y_2^n, S^n) + n(\epsilon_{1,n} + \epsilon_{2,n})$$

$$= I(M_1; Y_1^n | M_0, M_2, S^n) + I(M_0, M_2; Y_2^n | S^n) + n(\epsilon_{1,n} + \epsilon_{2,n})$$

$$= \sum_{t=1}^{n} I(M_1; Y_{1,t} | M_0, M_2, Y_1^{t-1}, S^n) + \sum_{t=1}^{n} I(M_0, M_2; Y_{2,t} | Y_{2,t+1}^n, S^n) + n(\epsilon_{1,n} + \epsilon_{2,n})$$

(A.25)

For the first term of (A.25) we have:

$$\sum_{t=1}^{n} I(M_1; Y_{1,t} | M_0, M_2, Y_1^{t-1}, S^n) \leq \sum_{t=1}^{n} I(M_1, Y_{2,t+1}^n; Y_{1,t} | M_0, M_2, Y_1^{t-1}, S^n)$$

$$= \sum_{t=1}^{n} I(Y_{2,t+1}^n; Y_{1,t} | M_0, M_2, Y_1^{t-1}, S^n) + \sum_{t=1}^{n} I(M_1; Y_{1,t} | M_0, M_2, Y_1^{t-1}, Y_{2,t+1}^n, S^n)$$

(A.26)

Also, for the second term of (A.25) we have:

$$\sum_{t=1}^{n} I(M_0, M_2; Y_{2,t} | Y_{2,t+1}^n, S^n) \leq \sum_{t=1}^{n} I(M_0, M_2, Y_{2,t+1}^n; Y_{2,t} | S^n)$$

$$= \sum_{t=1}^{n} I(M_0, M_2, Y_{2,t+1}^n, Y_1^{t-1}; Y_{2,t} | S^n) - \sum_{t=1}^{n} I(Y_1^{t-1}; Y_{2,t} | M_0, M_2, Y_{2,t+1}^n, S^n)$$

(A.27)

By combining (A.25)-(A.27), we obtain:

$$n(R_0 + R_1 + R_2) \leq \sum_{t=1}^{n} I(Y_{2,t+1}^n; Y_{1,t} | M_0, M_2, Y_1^{t-1}, S^n) + \sum_{t=1}^{n} I(M_1; Y_{1,t} | M_0, M_2, Y_1^{t-1}, Y_{2,t+1}^n, S^n)$$

$$+ \sum_{t=1}^{n} I(M_0, M_2, Y_{2,t+1}^n, Y_1^{t-1}; Y_{2,t} | S^n) - \sum_{t=1}^{n} I(Y_1^{t-1}; Y_{2,t} | M_0, M_2, Y_{2,t+1}^n, S^n) + n(\epsilon_{1,n} + \epsilon_{2,n})$$

(A.28)

One the one hand, based on the the Csiszar-Korner lemma we have:

$$\sum_{t=1}^{n} I(Y_{2,t+1}^n; Y_{1,t} | M_0, M_2, Y_1^{t-1}, S^n) = \sum_{t=1}^{n} I(Y_1^{t-1}; Y_{2,t} | M_0, M_2, Y_{2,t+1}^n, S^n)$$

(A.29)

Now, by substituting (A.29) in (A.28) we derive:

$$n(R_0 + R_1 + R_2) \leq \sum_{t=1}^{n} I(M_1; Y_{1,t} | M_0, M_2, Y_1^{t-1}, Y_{2,t+1}^n, S^n) + \sum_{t=1}^{n} I(M_0, M_2, Y_{2,t+1}^n, Y_1^{t-1}; Y_{2,t} | S^n) + n(\epsilon_{1,n} + \epsilon_{2,n})$$

$$= \sum_{t=1}^{n} I(X_t; Y_{1,t} | U_t, S_t) + \sum_{t=1}^{n} I(M_0, M_2, Y_{2,t+1}^n, Y_1^{t-1}; Y_{2,t} | S^n) + n(\epsilon_{1,n} + \epsilon_{2,n})$$

$$\leq \sum_{t=1}^{n} I(X_t; Y_{1,t}|U_t, S_t) + \sum_{t=1}^{n} I(M_0, M_2, Y_{2,t+1}^n, Y_1^{t-1}, S^{t-1}, S_{t+1}^n; Y_{2,t}|S_t) + n(\epsilon_{1,n} + \epsilon_{2,n})$$

$$= \sum_{t=1}^{n} I(X_t; Y_{1,t}|U_t, S_t) + I(U_t; Y_{2,t}|S_t) + n(\epsilon_{1,n} + \epsilon_{2,n})$$

(A.30)

Again, for the sum rate we have:

$$n(R_0 + R_1 + R_2) \leq I(M_0, M_1; Y_1^n, S^n) + I(M_2; Y_2^n, M_0, M_1, S^n) + n(\epsilon_{1,n} + \epsilon_{2,n})$$

$$= I(M_0, M_1; Y_1^n|S^n) + I(M_2; Y_2^n|M_0, M_1, S^n) + n(\epsilon_{1,n} + \epsilon_{2,n})$$

$$= \sum_{t=1}^{n} I(M_0, M_1; Y_{1,t}|Y_1^{t-1}, S^n) + \sum_{t=1}^{n} I(M_2; Y_{2,t}|M_0, M_1, Y_{2,t+1}^n, S^n) + n(\epsilon_{1,n} + \epsilon_{2,n})$$

(A.31)

For the first term of (A.31), we have:

$$\sum_{t=1}^{n} I(M_0, M_1; Y_{1,t}|Y_1^{t-1}, S^n) \leq \sum_{t=1}^{n} I(M_0, M_1, Y_1^{t-1}; Y_{1,t}|S^n)$$

$$= \sum_{t=1}^{n} I(M_0, M_1, Y_1^{t-1}, Y_{2,t+1}^n; Y_{1,t}|S^n) - \sum_{t=1}^{n} I(Y_{2,t+1}^n; Y_{1,t}|M_0, M_1, Y_1^{t-1}, S^n)$$

(A.32)

Also, for the second term of (A.31), we have:

$$\sum_{t=1}^{n} I(M_2; Y_{2,t}|M_0, M_1, Y_{2,t+1}^n, S^n) \leq \sum_{t=1}^{n} I(M_2, Y_1^{t-1}; Y_{2,t}|M_0, M_1, Y_{2,t+1}^n, S^n)$$

$$= \sum_{t=1}^{n} I(Y_1^{t-1}; Y_{2,t}|M_0, M_1, Y_{2,t+1}^n, S^n) + \sum_{t=1}^{n} I(M_2; Y_{2,t}|M_0, M_1, Y_1^{t-1}, Y_{2,t+1}^n, S^n)$$

(A.33)

Then, by combining (A.32) and (A.33), we derive:

$$n(R_0 + R_1 + R_2) \leq \sum_{t=1}^{n} I(M_0, M_1, Y_1^{t-1}, Y_{2,t+1}^n; Y_{1,t}|S^n) - \sum_{t=1}^{n} I(Y_{2,t+1}^n; Y_{1,t}|M_0, M_1, Y_1^{t-1}, S^n)$$

$$+ \sum_{t=1}^{n} I(Y_1^{t-1}; Y_{2,t}|M_0, M_1, Y_{2,t+1}^n, S^n) + \sum_{t=1}^{n} I(M_2; Y_{2,t}|M_0, M_1, Y_1^{t-1}, Y_{2,t+1}^n, S^n) + n(\epsilon_{1,n} + \epsilon_{2,n})$$

(A.34)

$$\stackrel{(a)}{=} \sum_{t=1}^{n} I(M_0, M_1, Y_1^{t-1}, Y_{2,t+1}^n; Y_{1,t}|S^n) + \sum_{t=1}^{n} I(M_2; Y_{2,t}|M_0, M_1, Y_1^{t-1}, Y_{2,t+1}^n, S^n) + n(\epsilon_{1,n} + \epsilon_{2,n})$$

$$\leq \sum_{t=1}^{n} I(M_0, M_1, Y_1^{t-1}, Y_{2,t+1}^n, S^{t-1}, S_{t+1}^n; Y_{1,t}|S_t) + \sum_{t=1}^{n} I(X_t; Y_{2,t}|V_t, S_t) + n(\epsilon_{1,n} + \epsilon_{2,n})$$

$$= \sum_{t=1}^{n} I(V_t; Y_{1,t}|S_t) + \sum_{t=1}^{n} I(X_t; Y_{2,t}|V_t, S_t) + n(\epsilon_{1,n} + \epsilon_{2,n})$$

(A.35)

where equality (a) is due to Csiszar-Korner lemma. Finally, by applying a time-sharing argument, we obtain (14). ∎

> ***Proof of Theorem 6:***

The achievability of (19) can be obtained from (10) by substituting $W = U$ and $V = Y_1$, (when the channel is degraded, the resulting rate region from this substitution includes (19) as a subset). Furthermore, the achievability of (19) can be directly proved using the superposition coding similar to [4]. For the converse part, consider a length-$n$ code with vanishing error probability for the channel.

Define the RVs $U_t \triangleq (M_0, M_2, Y_2^{t-1}, S_{t+1}^n)$, $t = 1, \ldots, n$. The second bound of (19) is derived by following the same lines as [2]. To derive the first bound, we can write:

$$nR_1 \leq I(M_1; Y_1^n, Y_2^n, S^n | M_0, M_2) + n\epsilon_{1,n}$$

$$= I(M_1; Y_1^n, Y_2^n | M_0, M_2, S^n) + n\epsilon_{1,n}$$

$$= \sum_{t=1}^{n} I(X_t; Y_{1,t}, Y_{2,t} | M_0, M_2, S^n, Y_1^{t-1}, Y_2^{t-1}) + n\epsilon_{1,n}$$

$$\stackrel{(a)}{=} \sum_{t=1}^{n} I(X_t; Y_{1,t} | M_0, M_2, S^n, Y_1^{t-1}, Y_2^{t-1}) + n\epsilon_{1,n}$$

$$= \sum_{t=1}^{n} H(Y_{1,t} | M_0, M_2, S^n, Y_1^{t-1}, Y_2^{t-1}) + n\epsilon_{1,n}$$

$$\stackrel{(b)}{\leq} \sum_{t=1}^{n} H(Y_{1,t} | U_t, S_t) + n\epsilon_{1,n}$$

$$(A.36)$$

where $\epsilon_{1,n} \to 0$ as $n \to \infty$, equality (a) holds because the channel is degraded and inequality (b) holds because conditioning does not increase the entropy. On the one hand, the presented proof for the converse part continues to hold for the case where CSI is also available at the stronger receiver, i.e., the first receiver. ∎

➢ *Proof of Theorem 8:*

Note that the scenario considered in this theorem can be derived from that one of Theorem 7 by replacing $S_1$ with $S_d$, $S_2$ with $S_r$, and $Y$ with $(Y, S_r)$. Therefore, the following rate is achievable:

$$\sup \min \begin{cases} I(U, U_r; Y, S_r) - I(U, U_r; S_r, S_d), \\ I(U; Y_r | U_r, S_d) - I(U; S_r | U_r, S_d) \end{cases}$$

$$(A.37)$$

where the supremum in $(A.37)$ is taken over all joint PDFs $P_{X_r U_r X U | S_r S_d} = P_{X_r U_r | S_d} P_{X U | U_r S_r S_d}$. Let $\alpha \in [0,1]$, and $X_r \sim \mathcal{N}(0, P_r)$ and $X_0 \sim \mathcal{N}(0, \alpha P)$ be two independent Gaussian RVs which are also independent of $(S_r, S_d)$. Define:

$$\begin{cases} X \triangleq \sqrt{\dfrac{\bar{\alpha} P}{P_r}} X_r + X_0 \\ U_r \triangleq \beta_r S_d + X_r \\ U \triangleq \beta_1 S_r + \beta_2 S_d + \beta_3 X_r + X \end{cases}$$

$$(A.38)$$

where,

$$\begin{cases} \beta_r = \dfrac{P_r + \sqrt{\bar{\alpha} P P_r}}{P + P_r + 2\sqrt{\bar{\alpha} P P_r} + N_r + N_d} \\ \beta_1 = \dfrac{\alpha P}{\alpha P + N_r} \\ \beta_2 = \dfrac{\alpha P}{\alpha P + N_r + N_d} \\ \beta_3 = \beta_2 \left( \sqrt{\dfrac{\bar{\alpha} P}{P_r}} + 1 \right) - \sqrt{\dfrac{\bar{\alpha} P}{P_r}} \end{cases}$$

$$(A.39)$$

By substituting $(A.38)$ in $(A.37)$, we derive the rate (28). ∎